\documentclass[proof]{WileyASNA-v1}

\usepackage[T1]{fontenc}
\usepackage{graphicx}
\usepackage{subfig}
\usepackage{rotating}

\articletype{ORIGINAL ARTICLE}%

\received{28 December 2018}
\revised{05 January 2019}
\accepted{18 January 2019}

\begin{document}

\title{An analysis of four stellar rings}

\author[1]{E. Paunzen*}

\author[2]{J. Florian}

\author[2]{A. G{\"u}tl-Wallner}

\author[2]{A. Herdin}

\author[2]{E. Kralofsky}

\author[2]{K. Leschinski}

\author[2]{M. Mach}

\author[2]{H.M. Maitzen}

\author[1]{M. Pri{\v s}egen}

\author[2]{M. Rockenbauer}

\author[4]{M. Rode-Paunzen}

\author[2,3]{S. Wallner}

\authormark{PAUNZEN \textsc{et al}}

\address[1]{\orgdiv{Department of Theoretical Physics and Astrophysics}, \orgname{Masaryk University}, \orgaddress{\state{Brno}, \country{Czech Republic}}}

\address[2]{\orgdiv{Institute of Astronomy}, \orgname{University of Vienna}, \orgaddress{\state{Vienna}, \country{Austria}}}

\address[3]{\orgdiv{ICA}, \orgname{Slovak Academy of Sciences}, \orgaddress{\state{Bratislava}, \country{Slovak Republic}}}

\address[4]{\orgdiv{BAS:IS (Library, Archiv, Collections)}, \orgname{Austrian Academy of Sciences}, \orgaddress{\state{Vienna}, \country{Austria}}}

\corres{*Department of Theoretical Physics and Astrophysics, Masaryk University, Kotl\'a\v{r}sk\'a 2, 611 37 Brno, Czech Republic. \email{epaunzen@physics.muni.cz}}

\abstract{About 50 years ago, one thousand ring-like structures (called stellar-rings) were discovered by \citet{Isser68}.
They were believed to be groups of young stars formed by shell-like triggered star formation which would make them
excellent tracers of spiral arms, for example. Neglected for 40 years, we used highly accurate kinematic, astrometric, 
and photometric data in order to investigate the four most prominent stellar rings. The aim is to investigate if those 
structures are indeed physically related groups of stars. We used proper motions and parallaxes from the Gaia DR2 to 
calculate distances and to search for common properties. Colour-magnitude diagrams using $BVJHK_{\mathrm s}$ measurements 
were investigated and isochrones fitted. None of the four stellar rings consists of a physically related group of young stars. 
The location of stars in the line-of-sight mimics a ring-like structure on the sky. The colour-magnitude diagrams are typical 
for an integrated field population and not for a young star cluster, for example. 
The currently available data are sufficient to analyse ring-like structures with high statistical significance. This allows
a new search for such structures in the Milky Way.}


%\jnlcitation{\cname{%
%\author{Paunzen E.}, 
%\author{J. Florian},
%\author{A. G{\"u}tl-Wallner},
%\author{A. Herdin},
%\author{E. Kralofsky},
%\author{K. Leschinski},
%\author{M. Mach},
%\author{H.M. Maitzen},
%\author{M. Pri{\v s}egen},
%\author{M. Rockenbauer},
%\author{M. Rode-Paunzen}, and
%\author{S. Wallner}} (\cyear{2019}),
%\ctitle{An analysis of four stellar rings}, \cjournal{Astron.Nachr./AN}, \cvol{2019}.}

%%\fundingInfo{Funding info text.}

\maketitle

\footnotetext{\textbf{Abbreviations:} HSOY, Hot Stuff for One Year; POSS, Palomar Observatory Sky Survey; UCAC5, US Naval Observatory CCD Astrograph Catalog}

\section{Introduction} \label{introduction}

Fifty years ago, \citet{Isser68} introduced the so-called stellar rings,
as shell type stellar aggregates which in projection on the sphere appear as rings.
In the following ten years, this topic attracted huge attention. The main conclusion 
was that these rings consist of young stars in ellipsoidal configurations,
having a sharp outer boundary and constant minor diameter. \citet{Isser70}
showed that they are excellent tracers of the Galactic spiral arms because of their
youth.

The idea about the formation of stellar rings is based on triggered star formation by 
supernovae or other bubble-producing processes \citep{Lindb67}.
The large-scale structure of the cold interstellar medium (ISM) can be
significantly affected by violent events. This structure, which
is diverse with the complex distribution of shells, cavities, filaments,
arcs, and loops, is often referred to as the ``Cosmic
Bubble Bath'' \citep{Brand75}. The evolution of bubbles
produced by Supernovae (SN) explosions and stellar
winds of associations, are the primary processes that determine
the structure and energetics of all components of the diffuse
ISM. Studying SN explosions in
a uniform medium, \citet{Cox74} pointed out, that if the
Galactic SN rate is high enough, it can produce the ``Swiss cheese
morphology'' of the cold diffuse ISM with hot coronal gas inside
the bubbles. The model by \citet{McKee77} describes how SN explosions in the cloudy ISM
produce a three phase medium. In this scenario, the next generation of stars are
born in the compressed medium of the bubble walls.
This led to the search for so-called far-infrared loops \citep{Kiss04} and interstellar 
bubbles \citep{Beau14}.

Are stellar rings manifestations of such a violent and turbulent star formation?
If so, the members of such rings have to be at the same distance from the Sun, have a common mean proper motion, and have
the same age within the time scale of star formation (a few million years). 
Ever since the paper by \citet{Vidal73}, the nature of the most prominent rings has never been
analysed in more detail, probably because of the lack of precise kinematic and astrometric 
data.

In this paper, we analyse the properties of four different stellar rings (Aquila, Hydra, Orion, and
S58) which have been investigated in the past. For the stars within these rings, spectral information,
photometry, kinematic and astrometric data are available. We construct the colour-magnitude diagrams
using $BVJHK_{\rm S}$ photometry, derive distances, and use proper motions from the Gaia DR2 
\citep{Linde18} data. The latter were also compared with those from the US Naval Observatory 
CCD Astrograph Catalog \citep[UCAC5,][]{Zacha17} and 
Hot Stuff for One Year \citep[HSOY,][]{Altm17} to search for possible off-sets.

\begin{center}
\begin{table}[t]%
\caption{The four investigated stellar rings together with the equatorial and Galactic coordinates
of the centres as well as the number of ``members'' (see text for the definition).}
\label{coords}
\centering
{\footnotesize
\begin{tabular}{@{\extracolsep\fill}lccccc@{\extracolsep\fill}}
\toprule
Name & $\alpha$ (2000) & $\delta$ (2000) & $l$ & $b$ & $N$ \\
\midrule
Aquila	& 20:05:18.2	& +11:24:16.9	& 51.789	& $-$10.729 & 33 \\
Hydra	& 10:17:00.5	& $-$17:19:58.2	& 258.392 & +31.873	& 27	\\
Orion	& 05:35:24.7	& $-$01:15:01.2	& 205.158	& $-$17.442 & 26 \\
S58	& 02:21:58.7	& +59:11:23.4	& 134.306	& $-$1.668 & 45	\\
\bottomrule
\end{tabular}
}
\end{table}
\end{center}

\section{Target selection and data sources} \label{selection}

We have selected the four most investigated and best defined stellar rings \citep{Isser75} in order to analyse their characteristics.
The names and coordinates are listed in Table \ref{coords}. In the following, we give an overview of the already
published results of them.

{\it Aquila:} \citet{Isser69} obtained photoelectric $UBV$ measurements of stars of the ring and concluded that the
minor diameter is 7.4\,pc, the distance from the Sun about 250\,pc and the age of 4\,Myr. The first analysis of the
individual proper motions was done by \citet{Proch71}. Due to the errors of the proper motions
between 0.3'' and 2.0'', no certain conclusion was drawn, but the intrinsic dispersion of the ring members
was significantly larger than predicted by \citet{Isser68}. Later on, \citet{Webst72} presented new spectral types
and concluded that the color-magnitude diagram looks not unlike that of a sparse Galactic cluster. But the distribution
of stars differs only one standard deviation from that of a suitably chosen comparison field. \citet{Isser74} 
concluded that the UV-excess (on the basis of MK-classification and photoelectric $UBV$ data) of many stars of this ring 
are hard to reconcile with a chance configuration of field stars. The large scatter of distance moduli reported before,
is due to the wrong usage of the calibration based on the UV-excesses.

{\it Hydra:} \citet{Schild71} derived spectral types of 26 stars of this stellar ring. The most intriguing fact is
that 16 of these stars have a luminosity class III, several of them being of K- and M-type which is clearly inconsistent 
with the hypothesis of a young age for the stellar rings (Sect. \ref{character}). Furthermore, the Hertzsprung-Russell diagram showed many 
inconsistencies, such as reddened F-type stars.
The mean distance modulus is given as 6.5$\pm$1.3\,mag, which corresponds
to an uncertainty of the distance of about 150\,pc. Overall, they concluded that there is a lack of physical
connection for this stellar ring.

{\it Orion:} \citet{Schild71} published spectral types of 27 stars which are consistent with the age of the
Orion complex. But they also stated that their results do not bear on the question if it is observationally
distinguishable from the remaining stars of the Orion belt. 

{\it S58:} \citet{Koles73} determined the spectrophotometric distances of 26 stars and the colour-magnitude diagram
of 720 stars in the surrounding of this stellar ring. For these 26 stars, they found a spread between 70 and 2630\,pc
and concluded that the ring is an accidental projection of stars on the celestial sphere, which are located
at various distances. 

In the following we define the terminus ``member'' of an individual stellar ring, as a star which is part of
it, on the basis of the original findings by \citet{Isser68} from the POSS plates. 
In total, 134 stars were identified, from which 131 objects
have parallaxes and proper motions in the Gaia DR2. These stars were considered for further analysis.
To test the consistency of the Gaia DR2, proper motions were also taken from the HSOY, and UCAC5.

For generating the colour-magnitude diagrams, we have used the $BV$ data published by \citet{Kharc01}.
This all-sky catalogue of more than 2.5 million stars transformed the Hipparcos/Tycho $BV$ magnitudes in
standard Johnson ones, using a homogenized transformation law. The $JHK_{\mathrm s}$ magnitudes were taken from the
2MASS 6X Point Source Working Database \citep{Skrut06}. 

\begin{figure}[t]
\begin{center}
\includegraphics[width=85mm]{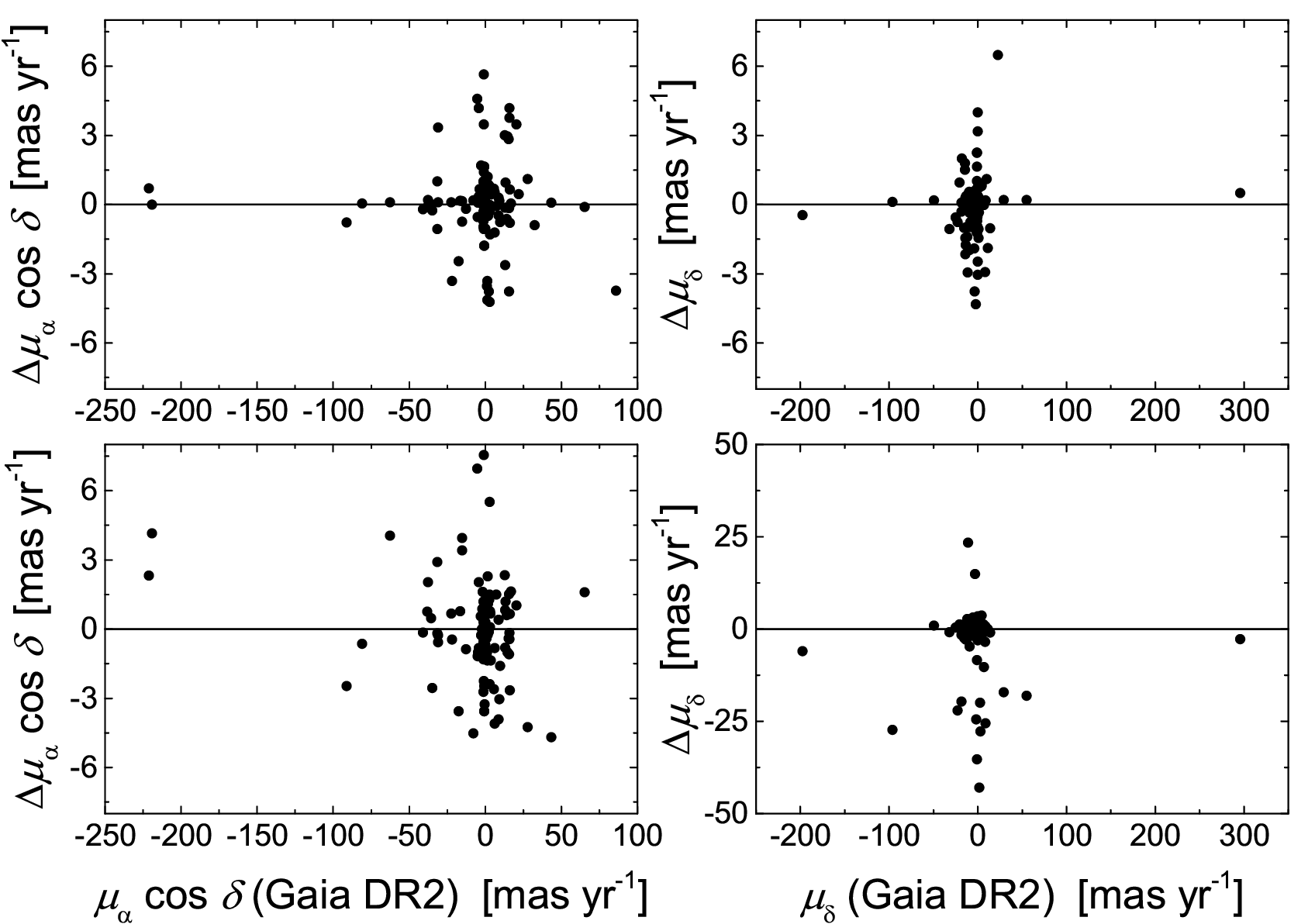}
\caption{The comparison of the proper motions for all stars of the four stellar rings from the HSOY (upper panels) and
the UCAC5 (lower panels) with the Gaia DR2. Note the different scaling for the lower 
right panel.}
\label{pm_comparison}
\end{center}
\end{figure}

\begin{figure*}[t]
\begin{center}
\includegraphics[width=170mm]{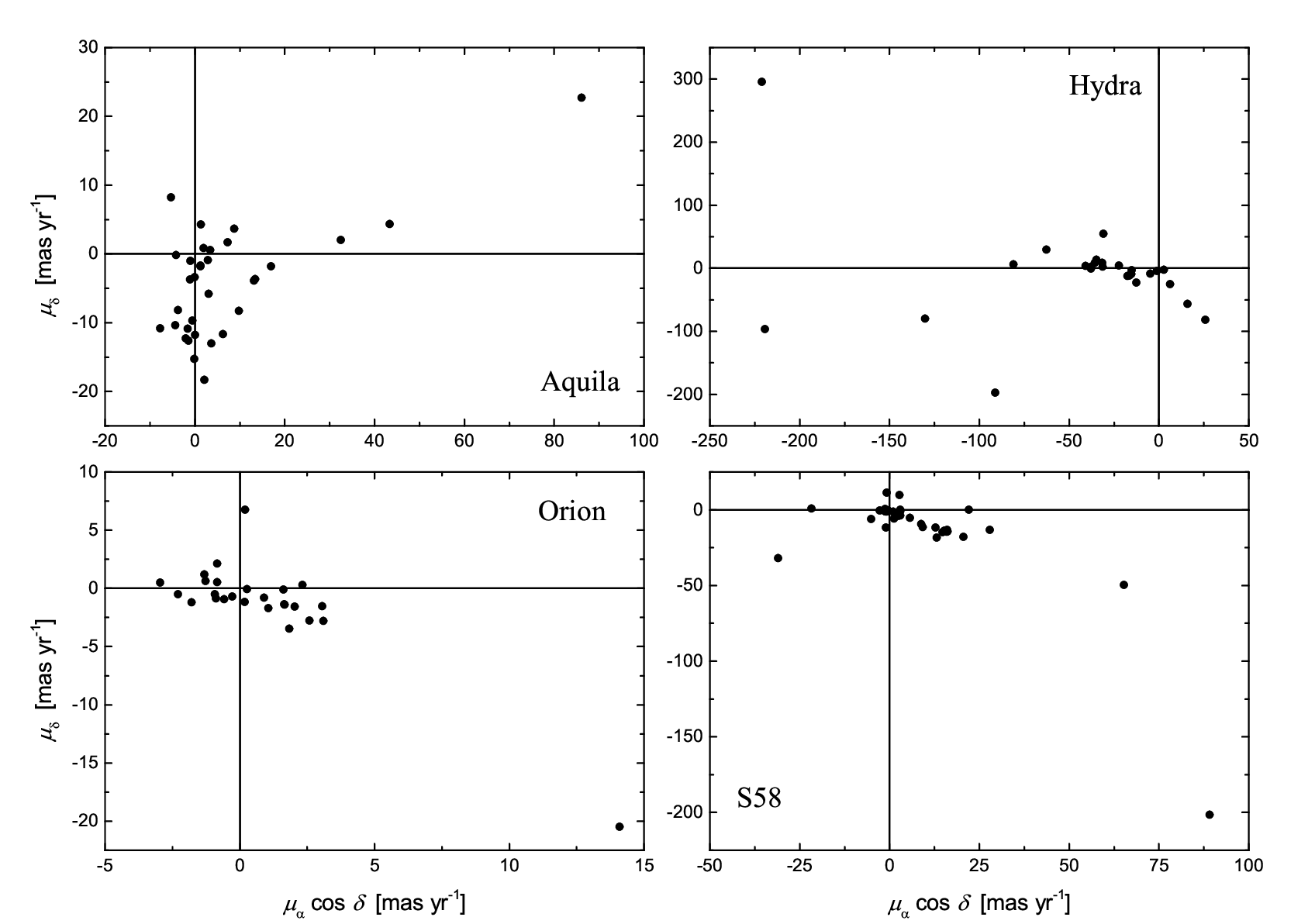}
\caption{The proper motions taken from Gaia DR2 of the four investigated
stellar rings. The errors of the individual values are smaller than the
symbol size.
The typical spread of the proper motions for members of 
open clusters and stellar moving groups is about $\pm$3\,mas\,yr$^{-1}$.}
\label{proper_motions}
\end{center}
\end{figure*}

The data of all 131 stars with their coordinates,
proper motions, parallaxes, and photometry are only available in electronic form. 

\section{Characteristics of stellar rings} \label{character}

Here we review the characteristics of stellar rings as described in a series of papers after the first
report by \citet{Isser68}. He was the first who published a paper on a new phenomenon called stellar rings. 
He detected 1002 ring-shaped groups of stars using the Palomar Observatory Sky Survey (POSS). 
They turned out to be even more numerous than stellar open clusters at the time of the publication. 
His working hypothesis was that these rings consist of stars at the same distance, i.e. they are somehow
gravitionally bound to each other.
Because these stellar rings are concentrated to the Galactic plane it was possible to determine, together with the 
help of POSS, some of the physical parameters. He especially tried to find the sizes of the rings that were smaller 
than the field of view of the observations, such that a complete ring was featured. 
If only fragments of a ring was visible, it was not included in the catalogue. 

He came to realize, that if there really are structures like stellar rings, they have to be very young with a high 
amount of stars of early spectral types because of dynamical reasons (e.g. the differential rotation of the Milky Way
would dissolve them very fast). His theory on the rings was supported by the 
fact that OB stars appear in those ring-shaped group of stars more frequently than among field stars in the vicinity. 
So stellar rings should not be just a chance configuration.
From this point-of-view, stellar rings are excellent tracers of the spiral arms, where star formation is still
ongoing. 

There are seven very important points concluded about stellar rings
\begin{itemize}
\item They consist of 25 to 200 stars
\item The thickness of the rings amounts from 1/10 to 1/40 of their semi-minor axes $D_{\rm m}$
\item The rings are nearly symmetrical relative to their semi-major axes $D_{\rm M}$
\item The ratio between both axes is always $D_{\rm M}$/$D_{\rm m}$\,$\le$\,2.0
\item The semi-minor axes are between 0.7' and 45'
\item The absolute mean semi-minor axis is about 7.1\,pc
\item Like for star clusters, the members should have a common mean proper motion with only a small deviation
\item The larger rings are always seen on the red and blue photographic plates, the smaller ones are hardly seen on the blue plates
\item Usually, there are more luminous stars in larger rings
\end{itemize}

\citet{Isser68} also developed a model about how these stellar rings are formed. 
From the size and the final shape of the rings it was deduced that they could have only been formed from separated dark clouds. 
The star formation from the interstellar matter must have started at nearly the same time because otherwise the stars born first 
would have ionized the hydrogen that would have drifted apart by expansion. The typical diameter of interstellar clouds is well
in the range of the mean diameter of the rings. Because there is an enhancement of the central regions of a cloud during the
stellar formation, the material has to be transported outside (expanding gas-shell) to trigger formation at the outskirts of it.
Because they are young, the stars did not have the time to drift apart and so they have to be near the place of their 
birth. 

The only inconsistent observational fact, at that time, was the existence of K and M dwarfs among the 
members of the stellar rings. Such stars could not have been reached the main sequence in a few Myrs
\citep{Iben65}. 

However, \citet{Vidal73} were the first to challenge the existence of stellar rings on a statistical basis.
They searched for all apparent stellar rings within a Galactic longitude range of 120$^{\rm o}$ to 150$^{\rm o}$.
Then they compared the small and large diameters of those 72 rings in common with \citet{Isser68}. In total, they
found 257 ring-like structures on the POSS plates within a Galactic latitude of $\pm$10$^{\rm o}$. If those rings are
young and physical, the spiral arms should be easily detectable as published by \citet{Isser68}. However, they were
not able to reproduce the later result, their stellar rings were uniformly distributed. From this, they concluded that
the stellar rings are chance configurations in the line-of-sight, or in some cases chance configurations embedded in young
stellar aggregates and that the spiral structure found by \citet{Isser68} was due to small sampling and/or selection
effects.

\begin{figure}[t]
\begin{center}
\includegraphics[width=85mm]{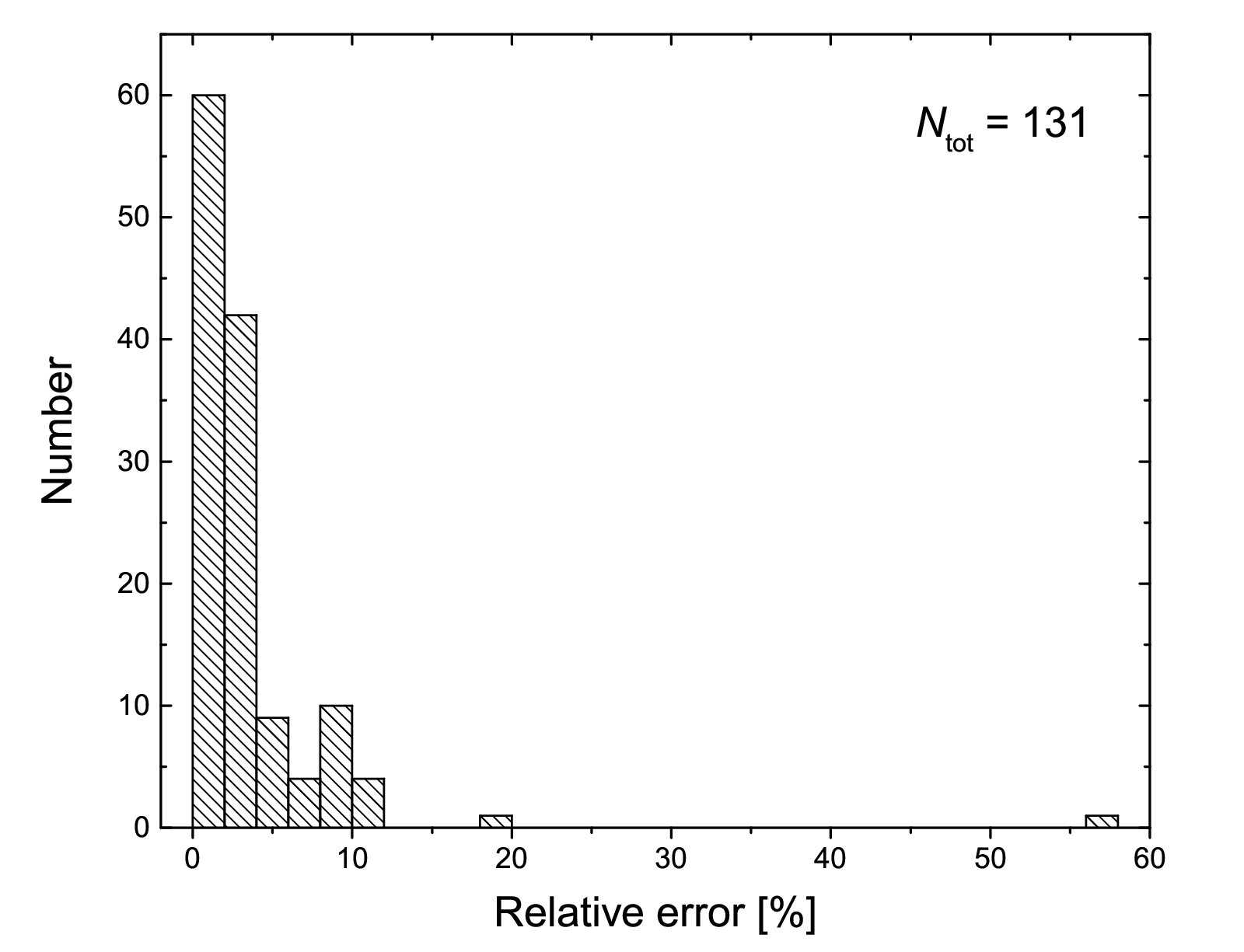}
\caption{The histogram of the relative errors of the Gaia DR2 parallaxes
for the 131 stars in the four stellar rings.}
\label{relative_errors}
\end{center}
\end{figure}

\begin{figure*}
   \centering
	 \begin{tabular}{cc}
   \subfloat[][Stellar ring in Aquila. \label{Aquila}]{\includegraphics[width=0.5\linewidth]{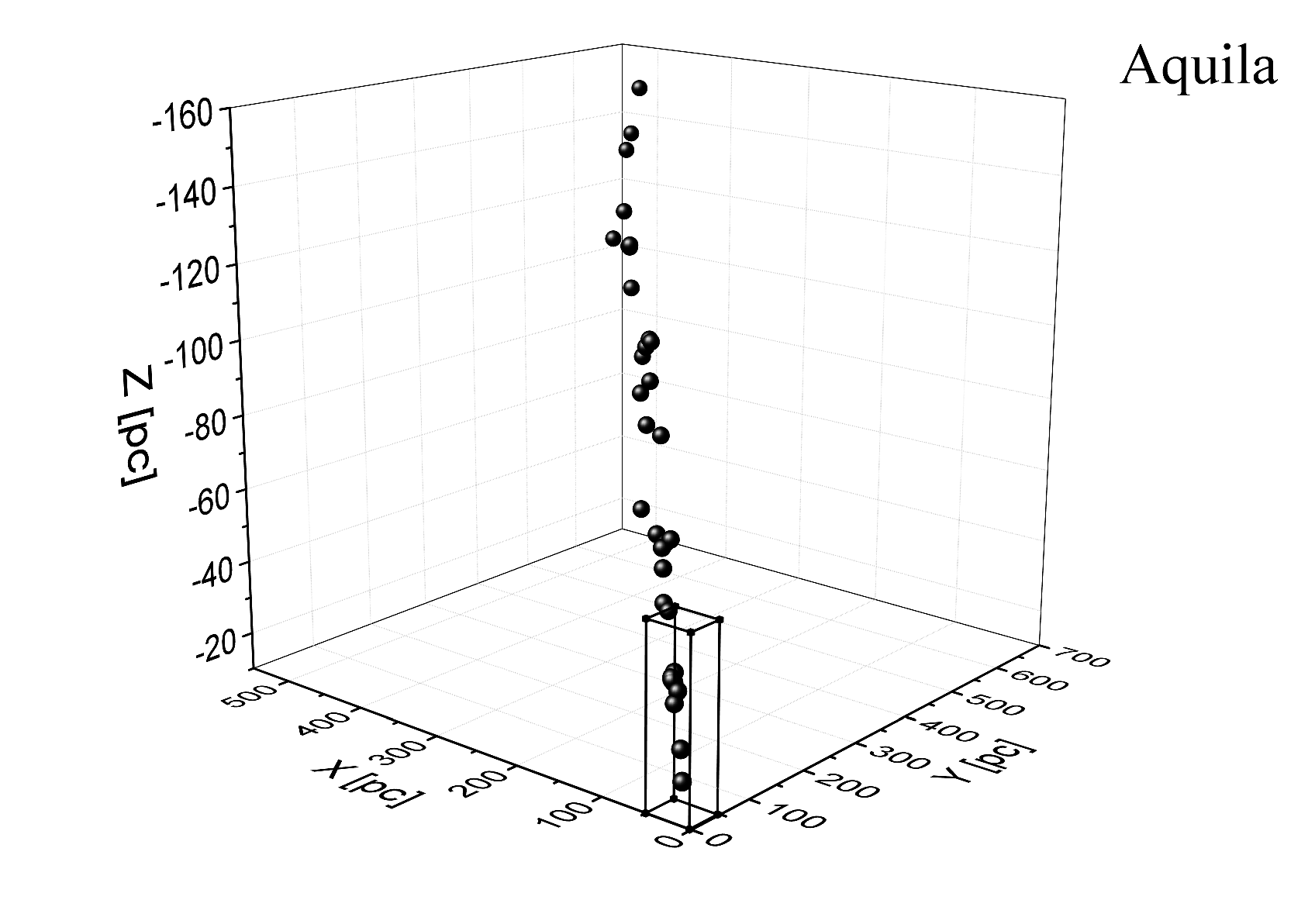}}\quad &
   \subfloat[][Stellar ring in Hydra. \label{Hydra}]{\includegraphics[width=0.5\linewidth]{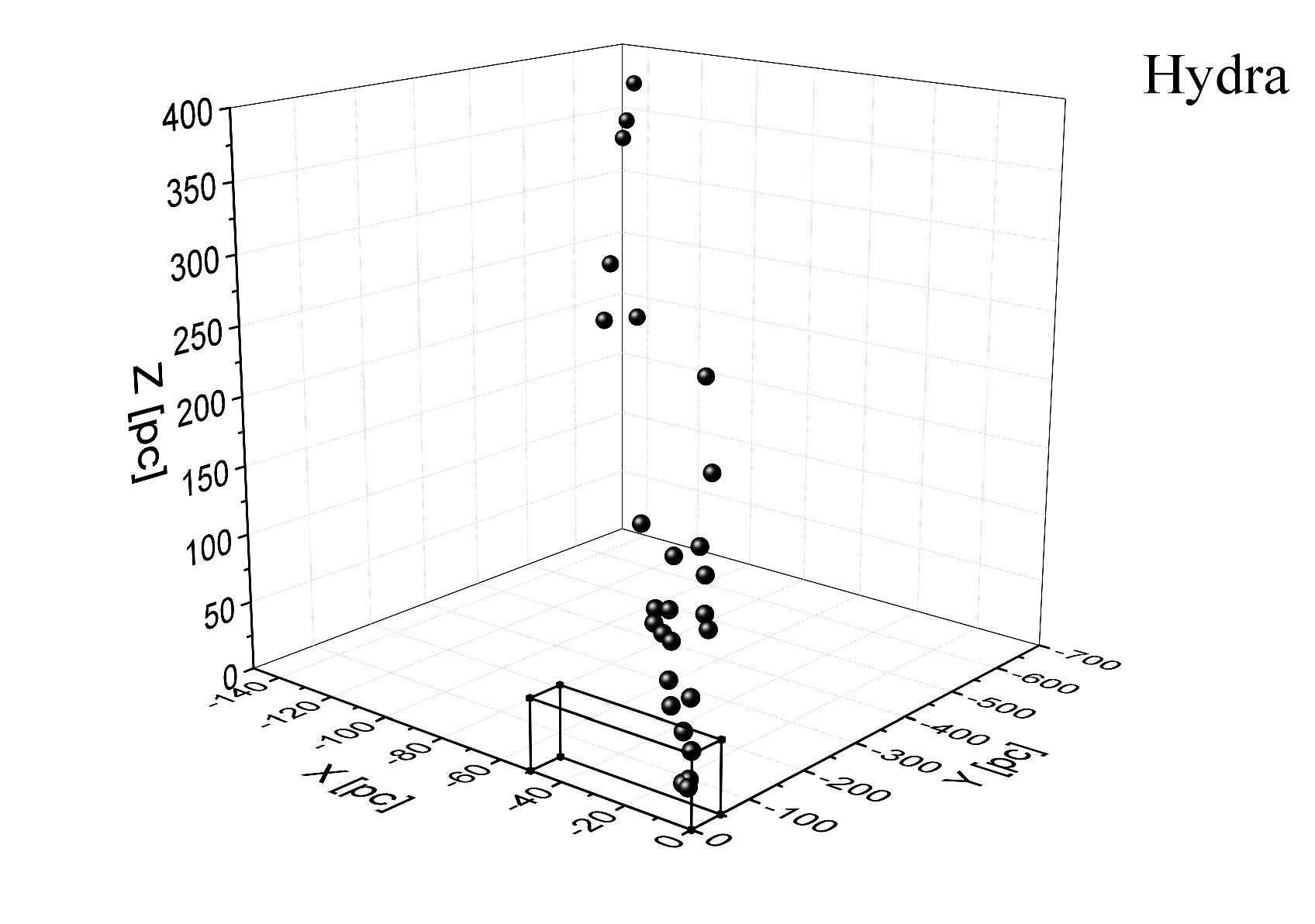}}\\
   \subfloat[][Stellar ring in Orion. \label{Orion}]{\includegraphics[width=0.5\linewidth]{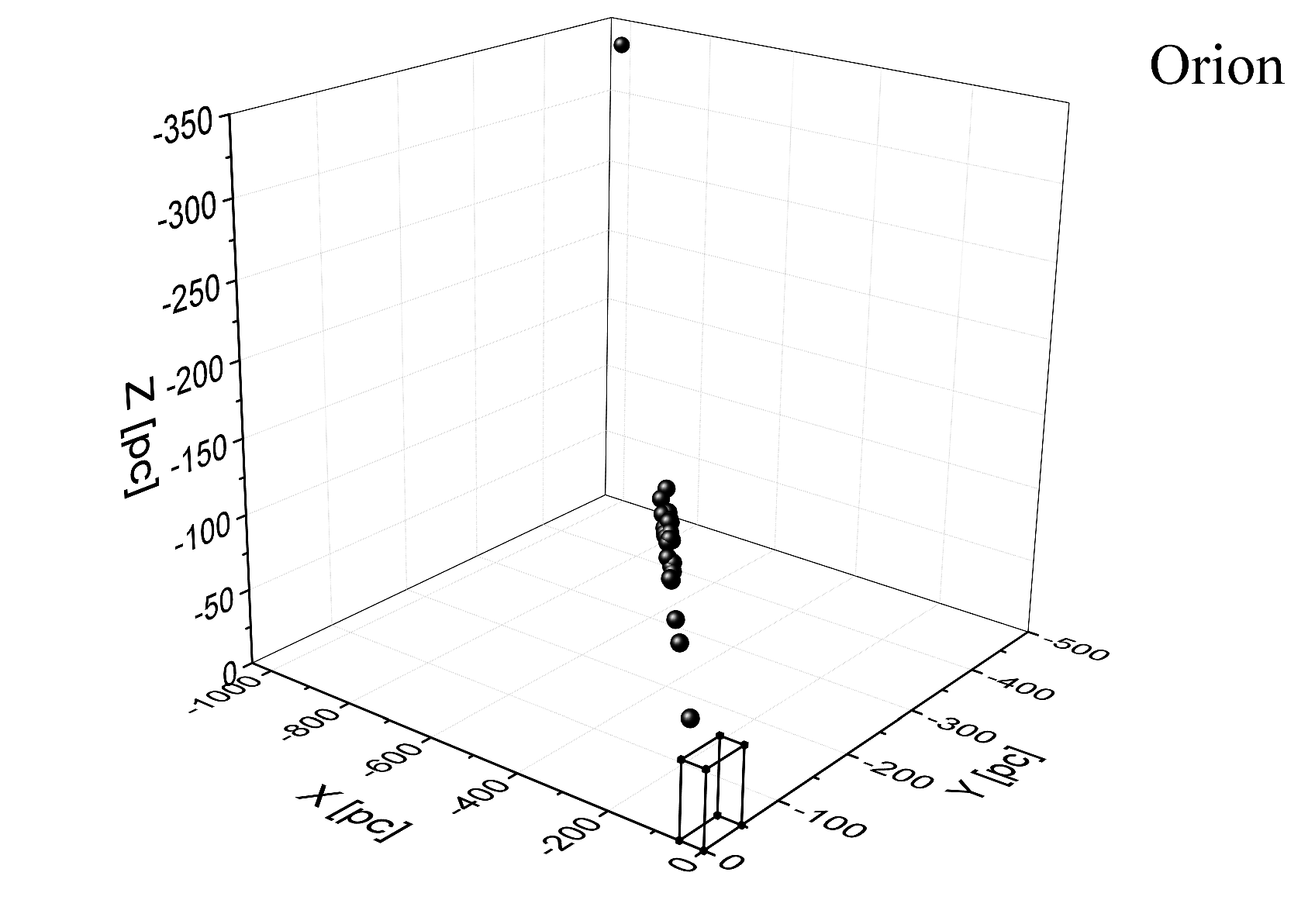}}\quad &
   \subfloat[][Stellar ring S58. \label{S58}]{\includegraphics[width=0.5\linewidth]{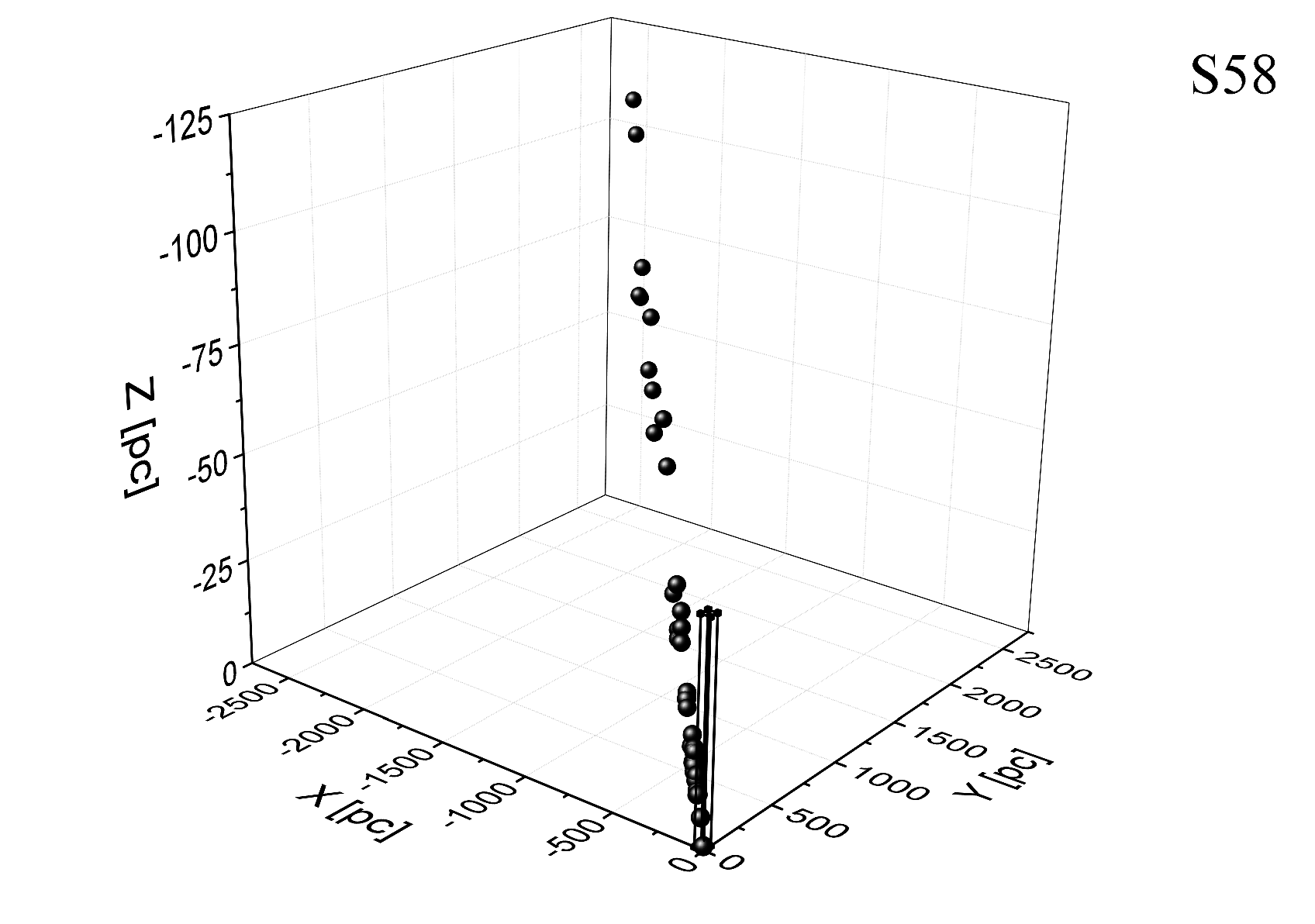}} \\
	 \end{tabular}
   \caption{The location of the proposed members of the four investigated stellar rings. The drawn boxes always have the
            size 50x50x50\,pc. The errors of the distances are smaller than the symbol size.}
   \label{positions}
\end{figure*}

\begin{figure}[ht]
\begin{center}
\includegraphics[width=85mm]{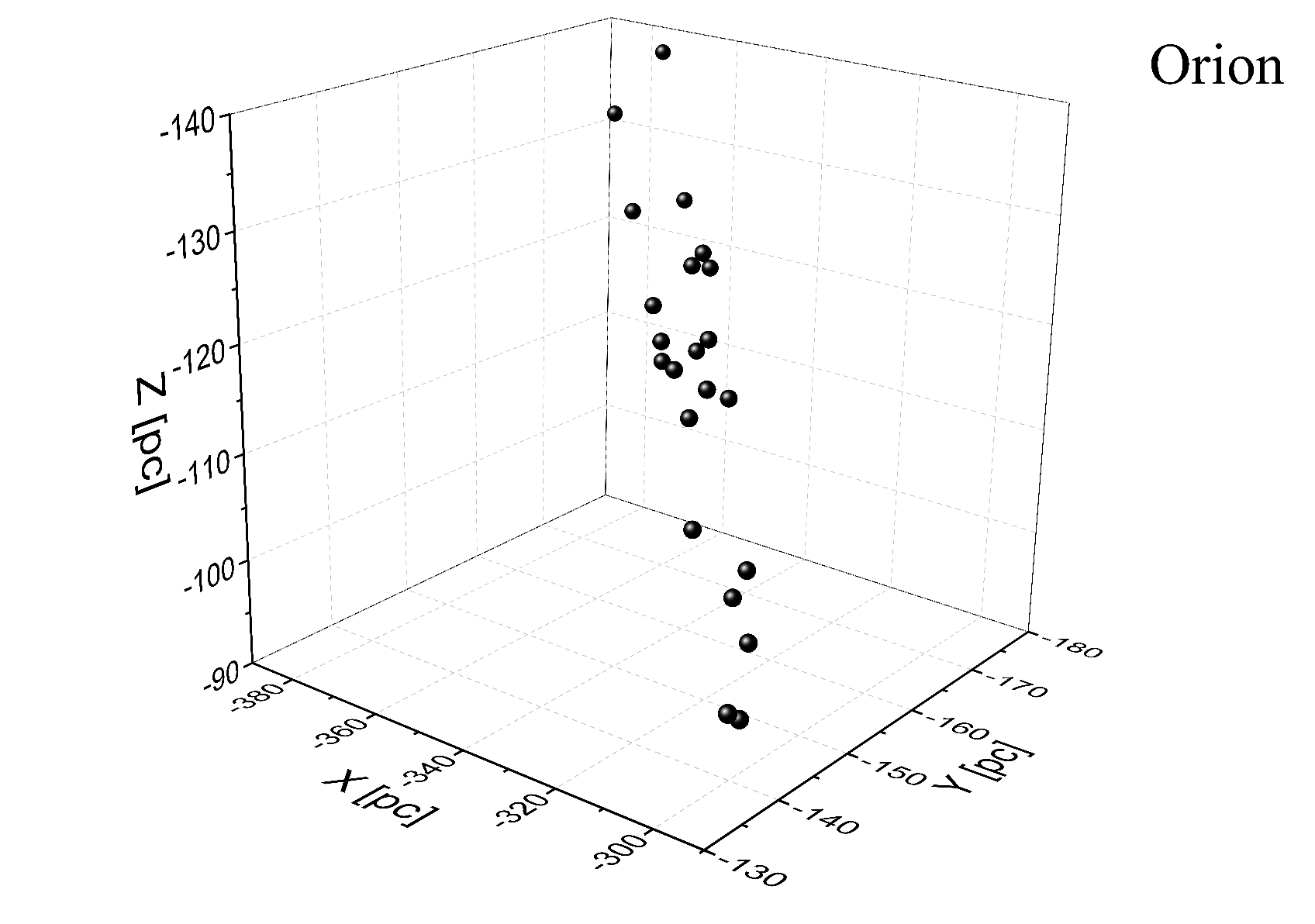}
\caption{An enlargement of the stellar ring in Orion. The region has a size of
	100x50x50\,pc. The errors of the distances are smaller than the
  symbol size.}
\label{Orion_zoom}
\end{center}
\end{figure}

\begin{figure*}[ht]
\begin{center}
\includegraphics[width=170mm]{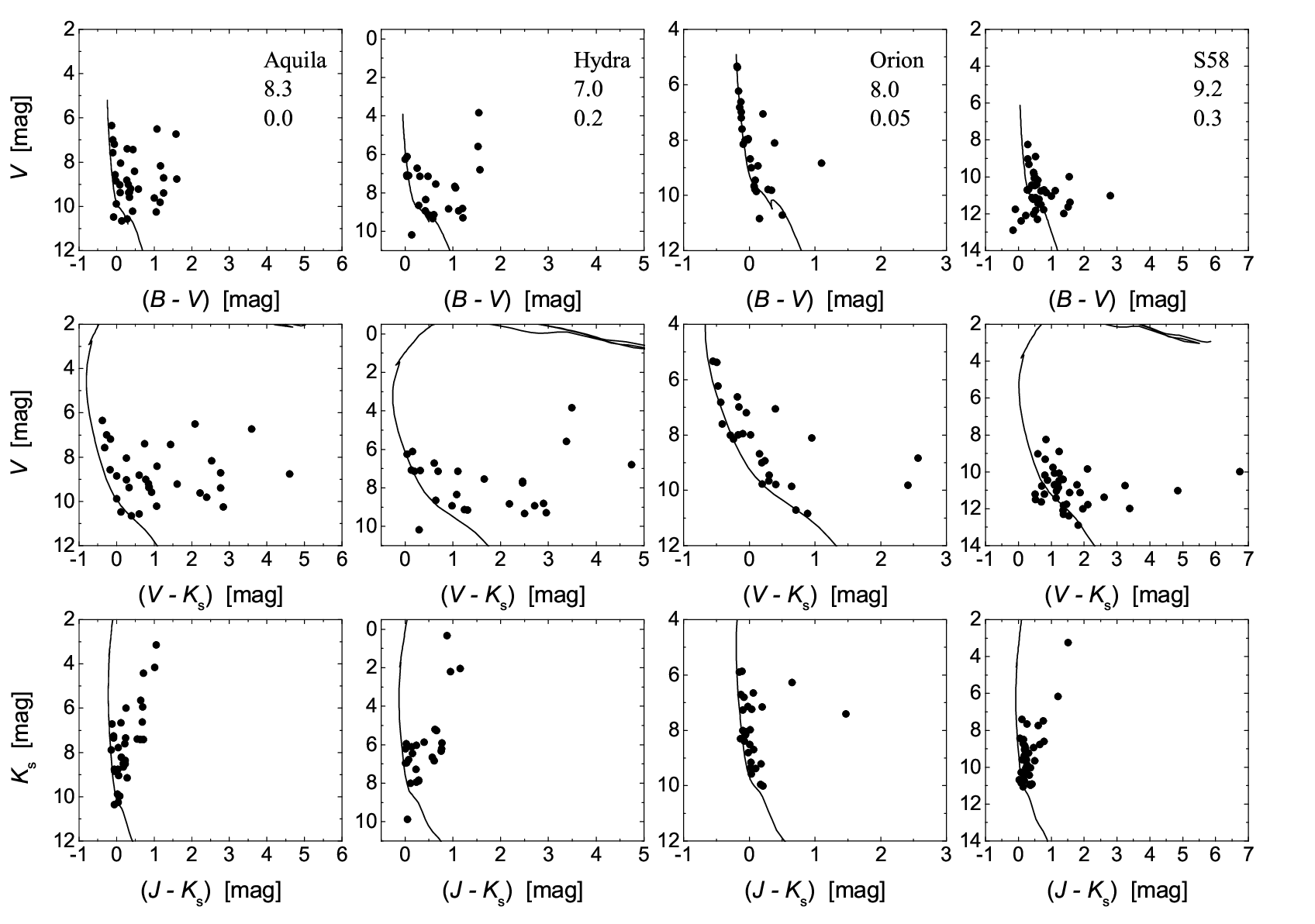}
\caption{The colour-magnitude diagrams together with the zero-age-main sequence (solar metallicity) taken from
\citet{Bressan12} for the four stellar rings. The scaling of the ordinate and
abscissa for each stellar ring is identical. The number below the name denote the absolute distance modulus
and the $E(B-V)$ value, both in units of magnitudes.}
\label{cmds}
\end{center}
\end{figure*}

\section{Analysis and results} \label{analysis}

First of all, we compared the proper motions from the HSOY \citep{Altm17} and 
UCAC5 \citep{Zacha17} with those from the Gaia DR2 to search for possible off-sets.
The HSOY and UCAC5 are based on independent ground-based observations and the Gaia DR1 \citep{Gaia2016}.
Therefore, we assume that they are of comparable quality and appropriate to identify 
inhomogeneities. The mean errors of the members of all four star rings in the HSOY are
$\pm$0.7\,mas\,yr$^{-1}$, and $\pm$1.7,mas\,yr$^{-1}$ for the UCAC5, respectively.
\citet{Fedor18} published an analysis of the HSOY and UCAC5 looking for possible 
offsets and correlations of the proper motions and the apparent magnitude of stars.
Within these two catalogues, they have found no such correlations. In Fig. \ref{pm_comparison}
the comparison of the proper motions from the HSOY (upper panels) and
the UCAC5 (lower panels) with the Gaia DR2 are shown. For the HSOY, we find a few stars which 
have differences larger than 3$\sigma$, but not exceeding $\pm$6.3\,mas\,yr$^{-1}$. Also the 
the $\mu_{\alpha} \cos \delta$-differences show the same behaviour. For the $\mu_{\delta}$-differences
of the UCAC5, we find 15 (or 14\%) stars with values larger than 15\,mas\,yr$^{-1}$. We can
exclude a wrong identification of these objects. Because there are no off-sets in the HSOY for
these stars, we conclude that the values in the UCAC5 are wrong. In overall, the Gaia DR2 and 
HSOY are highly consistent and suitable for our analysis.

The proper motions of our targets based on the Gaia DR2 are presented in Figure \ref{proper_motions}.
Let's keep in mind that
the typical spread of the proper motions for members of 
open clusters and stellar moving groups is about $\pm$3\,mas\,yr$^{-1}$ \citep{Dias18,Roeser18}.
This heuristically observed value is due to the separation of the primordial molecular cloud and the
conservation of angular momentum. From this figure we conclude that only the stellar ring in
Orion shows a spread which is compatible with the above given value. For the other rings, we mainly
see a group around the zero value which are the more distant background objects and the closer 
stars with large proper motions as well as an otherwise wide spread. Some more dense regions in
these diagrams are due to stars in certain distance ranges following the Galactic rotation \citep{Schoen12}.
Even if we take these stars as a group, they do not form a ring-like shape.

The Gaia DR2 is the first parallax data source allowing to study groups of stars such as 
stellar rings with high accuracy. Figure \ref{relative_errors} shows the relative errors
of the 131 stars in the four stellar rings. About 85\% of all objects have relative errors less than
5\% and even 95\% less than 10\%. As reviewed in Sect. \ref{selection}, divergent results about the
nature of the investigated targets were published in the past. The parallaxes and their errors were
directly converted into distances. To further study the location of the members of the individual
stellar ings, we have used the [$XYZ$] coordinates, i.e. towards the Galactic centre ($X$), the direction of 
the Galactic disk rotation ($Y$) and the North Galactic Pole ($Z$). Using the Galactic longitude ($l$) and latitude
($b$) as well as the distance from the Sun ($d$), the [$XYZ$] coordinates can be calculated as

\begin{eqnarray}
X&=&d \cos b \cos l,\\
Y&=&d \cos b \sin l,\\
Z&=&d \sin b.
\end{eqnarray}
Assuming that the error of the stars position is negligible, the error of the distance directly
transform to the three components. The results are shown in Figure \ref{positions}. In each panel we
included a box of the size 50x50x50\,pc, respectively. As originally defined (Sect. \ref{character}), the
absolute mean small diameter of stellar rings is about 7.1\,pc assuming that the members are all at the same
distance. From the figure, the projection effect of stars in one line-of-sight \citep{Meure79} is nicely
visible. The following distance ranges from the Sun were found

\begin{itemize}
\item Aquila: 77 to 823\,pc
\item Hydra: 33 to 758\,pc 
\item Orion: 154 to 1119\,pc
\item S58: 18 to 3283\,pc
\end{itemize}

For the Aquila and Hydra stellar rings, no groups with more than four stars in a volume of 10\,pc$^{3}$ and
a common proper motion were found.

Within S58, we noticed a group of nine stars at a distance between 350 and 390\,pc and a common
proper motion of [+15.3, $-$13.7] with a spread of about $\pm$1.2\,mas\,yr$^{-1}$. These parameters are in
coincidence with those from the ``Moving Group No. 36'' published by \citet{Oh17} on the basis of the Gaia DR1. 
It is interesting to note, that they have not included our nine stars in their list of members. Also \citet{Faher18}
who made a re-analysis of all groups on the basis of a much more extended data set, did not find new members.
This might be caused by the limited search radius of 10\,pc. We, therefore, looked in the newest version of
the catalogue by \citet{Dias2002} to search for an open cluster in this direction. Indeed, the open clusters Stock 2 is located at the corresponding
coordinates (02 14 43, +59 29 06) with a comparable proper motion (+17.29, $-$13.10). 
The age of Stock 2 is about 170\,Myr. However, the distance is
listed as 300\,pc by \citet{Dias2002} and was revised by \citet{Cantat2018} to 375\,pc. This shows 
the need of a homogeneous analysis of the Galactic open cluster population using the Gaia DR2 data and
forthcoming releases. As a summary it can be said that S58 is a mixture of a small group of stars which
are members of the known open clusters Stock 2 and further fore- and background stars. It is not a physical
connected stellar ring in the classical sense. 

The situation of the Orion ring needs a further discussion. Already in the past, this was always the most
promising candidate for a physically related group of stars. In Fig. \ref{Orion_zoom} an enlargement of it
with a size of 100x50x50\,pc is shown. There are two substructures visible. The lower one consists of six
stars, the upper one of twelve objects (excluding the upper most four objects). Let's recall that the 
mean proper motions of the Orion complex is only a few mas\,yr$^{-1}$ (Fig.\,\ref{proper_motions}) and the 
mean radial velocity of its members is about +30\,km\,s$^{-1}$ \citep{Cottle2018}. One interesting 
object is HD\,290606 with a very high proper motion (the entry in the lower right corner of the corresponding
panel of Fig.\,\ref{proper_motions}) and a radial velocity of $-$21.26(35)\,km\,s$^{-1}$ \citep{Sartoretti2018}.
If we follow the theory that a SN explosion might cause the triggering of stellar formation, one of the by-products
of such an event in a binary system could be runaway stars \citep{Blaauw1961}. Originally, this theory was
developed to explain the O- and B-type stars with spatial high velocities. Later on, this scenario was also extended
to low-mass stars \citep{Vickers2015}. The only spectral type found for HD\,290606 is K0 taken from the HD 
catalogue \citep{Cannon1949}. Therefore, we have investigated its photometry (no spectroscopic data are available 
in the archives) together with the parallax in order to verify the spectral type.
HD\,290606 is located in the outskirts of the Orion complex. From the reddening map
by \citet{Green2018}, we deduced an extinction $E(B-V)$\,=\,0.12(2)\,mag. This transforms to an absorption in $V$
of 0.66\,mag using the known non-standard reddening law towards Orion \citep{Costero1970}. The photometry of the
data sources listed in Sect. \ref{selection} results in colours consistent of a K3 to K4 star. Using the parallax from
the Gaia DR2, we get an absolute magnitude of +0.24(5)\,mag. This corresponds to a luminosity class III K-type giant
and not to a main sequence star \citep{Gould1992}. However, we are not able to, a-priori, rule out a pre-main sequence 
nature of this object. According to the absolute magnitude and thus the luminosity, it has to be in the very early
stages of its evolution \citep{Vorobyov2017}. This stage is characterized by a significant IR-excess \citep{Meyer1997}.
First, we were searching for this signature using the VOSA (VO Sed Analyzer) tool v6.0 \citep{Bayo2008}. It was applied 
to fit the Spectral Energy Distribution (SED) to the available photometry. The best fit was achieved for 
a $T_{\mathrm{eff}}$ of 4\,500\,K without any signs of an IR-excess up to 22$\mu$m. Also in the classical near-IR 
colour-colour diagrams \citep{Meyer1997}, no signs of an excess is visible. We therefore conclude that HD\,290606
is an early K-type giant and does not originate within the Orion complex. As a summary, we conclude that also the
Orion stellar ring does not comply with the originally defined characteristics of a young stellar aggregate of stars
born due to triggered stellar formation.

Finally, we also investigated the colour-magnitude diagrams of the four stellar rings which lead to the conclusion
that they are physically related groups of young stars in the past works. In Fig. \ref{cmds} we present these diagrams using the $(B-V)$,
$(V-K_{\mathrm s}$, and $(J-K_{\mathrm s})$ colours. For each field, we list the absolute distance modulus
and the $E(B-V)$ value which fit the zero-age-main sequence best. The colour ranges for the Aquila, Hydra, and S58 stellar
rings are between 5.5 and 8\,mag, respectively. This is at least three magnitudes larger than what is expected for a
typical young open cluster \citep{Platais2007}. 
There are also a few giants and stars far from the zero-age-main sequence clearly visible in the 
$K_{\mathrm s}$ versus $(J-K_{\mathrm s})$ diagrams, which is characteristic for a field population
\citep{Jose2008} or an evolved open clusters if treated as members.
As already discussed before,
stars of the Orion ring are almost all located within this region and therefore form much better defined 
colour-magnitude diagrams. Also the location of the giant HD\,290606 is clearly visible. This stellar ring is a 
good example how a group of stars located within about 100\,pc could mimic a ring-like structure due to a common 
reddening. 
To estimate the colour-magnitude diagrams of background/foreground stars
in the field-of-view of our targets, we used the theoretical Galactic model, TRILEGAL 1.6\footnote{http://stev.oapd.inaf.it/trilegal} 
described by \citet{Girardi2005}. It includes the populations of the thin and thick disk as well as the Galactic halo. We have simulated 
fields of the corresponding sizes with the central coordinates as listed in Table \ref{coords}.
We restricted the sample to the $V$ magnitudes as deduced from Figure \ref{cmds}. For none of the colours, we find any significant 
difference of the characteristics between the synthetic and observed diagrams. The only difference is the number of stars
because for the stellar rings, only a few stars are specifically selected from the whole field. 

\section{Conclusions}

The highly accurate kinematic and astrometric data of the Gaia DR2 allowed, for the first time, an analysis of the four
best investigated and most prominent stellar rings, 50 years after their discovery. They were thought to be 
groups of young stars located within 10\,pc, formed together, triggered by a shell-like expansion of material
within one or several molecular clouds. Since these events are believed to be quite frequent (for example, SN explosions)
on a global scale, such rings would be perfect tracers of the star formation throughout the Milky Way. 

From the available data, we were able to conclude that three stellar rings consist of stars in widely different
distances from the Sun mimicking a ring-link structure projected on the sky. The stellar ring in Orion is mostly composed
of members from this stellar association. But the absolute dimension and its three-dimensional configuration also
confirms that it is not a stellar ring.

Furthermore, we investigated the $BVJHK_{\mathrm s}$ colours of all targets in order to understand the misleading
colour-magnitude diagrams. It shows that at least three rings also include giants which do not justify fitting
a zero-age-main sequence and thus a young age. The colours show a much larger spread than expected for a young
stellar cluster or a star forming region. 

The original survey by \citet{Isser68} was based on the POSS, done by inspecting the plates by eye.
He concentrated his search on the Galactic plane which increases the chances to find merely a random group
of stars as shown in this paper. He also reported countless unresolved and fragments of rings which were
not included in his catalogue. With the currently available precise kinematic, astrometric and photometric
all-sky data, it is possible to search for ring-like structures using high-performance computer algorithms
such as pattern recognition techniques.

\section*{Acknowledgments}

This publication makes use of data products from the Two Micron All Sky Survey, which is a joint project of the 
University of Massachusetts and the Infrared Processing and Analysis Center/California Institute of Technology, 
funded by the National Aeronautics and Space Administration and the National Science Foundation.
This work presents results from the European Space Agency (ESA) space mission Gaia. Gaia data are being processed 
by the Gaia Data Processing and Analysis Consortium (DPAC). Funding for the DPAC is provided by national institutions, 
in particular the institutions participating in the Gaia MultiLateral Agreement (MLA). This paper is dedicated to 
G. Paunzen who died during its preparation.

\nocite{*}% Show all bib entries - both cited and uncited; comment this line to view only cited bib entries;
\bibliography{Sternringe-ASNA}%

\end{document}